\documentclass[aps,twocolumn,nofootinbib,preprintnumbers,eqsecnum]{revtex4-1}

\usepackage{bbm}
\usepackage{mathbbol}
\usepackage[mathscr]{eucal}
\usepackage{bm}

\usepackage{color}

\newcommand{\nb}[1]{\color{blue}}

\usepackage[
	  pagebackref=false,
	  colorlinks=true,
      linkcolor=blue,
      urlcolor=blue,
      filecolor=black,
      citecolor=red,
      pdfstartview=FitV,
      pdftitle={Effective action for magnetohydrodynamics},
        pdfauthor={},
        pdfsubject={},
        pdfkeywords={},
        pdfpagemode=None,
        bookmarksopen=true
      ]{hyperref}

\usepackage[normalem]{ulem}
\usepackage{amsmath}
\usepackage{enumerate}
\usepackage{amsfonts}
\usepackage{epsfig}
\usepackage{mathbbol}
\usepackage{comment}

\def\Tr{\mathop{\rm Tr}}

\newcommand\p{\ensuremath{\partial}}

\newcommand{\be}{\begin{equation}}
\newcommand{\ee}{\end{equation}}
\newcommand{\bea}{\begin{eqnarray}}
\newcommand{\eea}{\end{eqnarray}}
\newcommand{\bega}{\begin{gather}}
\newcommand{\eega}{\end{gather}}

\newcommand{\bi}{\begin{itemize}}
\newcommand{\ei}{\end{itemize}}
\newcommand{\ben}{\begin{enumerate}}
\newcommand{\een}{\end{enumerate}}
\newcommand{\bca}{\begin{cases}}
\newcommand{\eca}{\end{cases}}
\newcommand{\bln}{\begin{align}}
\newcommand{\eln}{\end{align}}
\newcommand{\bst}{\begin{split}}
\newcommand{\est}{\end{split}}
\def\ie{\begin{equation}\begin{aligned}}
\def\fe{\end{aligned}\end{equation}}
\newcommand{\bma}{\le(\begin{matrix}}
\newcommand{\ema}{\end{matrix}\ri)}

\newcommand\ep{\epsilon}

\newcommand\vep{\varepsilon}

\def\le{\left}
\def\ri{\right}

\begin{document}
\title{Effective field theory of magnetohydrodynamics from generalized global symmetries}
\preprint{EFI-18-18}
\author{Paolo Glorioso and Dam Thanh Son}
\affiliation{Kadanoff Center for Theoretical Physics,
University of Chicago, Chicago, IL 60637, USA}
\begin{abstract}

We introduce an effective action for non-dissipative
magnetohydrodynamics. A crucial guiding principle is the generalized
global symmetry of electrodynamics, which naturally leads to
introducing a ``dual photon'' as the degree of freedom responsible for
the electromagnetic component of the fluid. The formalism includes
additional degrees of freedom and symmetries which characterize the
hydrodynamic regime. By suitably enhancing one of the symmetries, the
theory becomes force-free electrodynamics. The symmetries furthermore
allow to systematize local and non-local conserved helicities. We also
discuss higher-derivative corrections.

\end{abstract}
\maketitle

\section{Introduction}
%The power of effective field theory is in that it provides a convenient simple language to describe the low-energy behavior of a generic many-body system. Additionally, effective field theory approaches often offer a geometrical perspective which leads to unveiling new structures.

Magnetohydrodynamics (MHD)~\cite{LL8} combines Navier-Stokes
hydrodynamics and Maxwell electrodynamics into a theory that can
describes the long-time, long-distance behavior of charged fluids.  It
is applicable in systems of sizes ranging from laboratory systems to
the whole Universe.  In the standard formulation, the MHD equations
describe the time evolution of the hydrodynamic variables (density,
fluid velocity, temperature) and the magnetic field.

Recently, MHD has been recast into an alternative form using a
generalized global symmetry \cite{Grozdanov:2016tdf} associated with
the conservation of the dual of the Maxwell field strength:\footnote{We choose the convention $\vep^{0123}=\frac 1{\sqrt{-g}}$.}
\be \label{cons0}\p_\mu \tilde F^{\mu\nu}=0,\qquad \tilde
F^{\mu\nu}=\frac 12\vep^{\mu\nu\alpha\beta}F_{\alpha\beta}\ .\ee In
this formulation, MHD is a hydrodynamic theory with a conserve charge,
but the conserved current in this case is a 2-form, rather than a
1-form \cite{Gaiotto:2014kfa}. Equation~(\ref{cons0}) is the statement
that the magnetic flux, defined as the integral of the magnetic field
over a codimension-2 surface, is conserved.  This approach has the
advantage that all equations of MHD are treated in the same footings,
i.e., as conservation equations.

In this paper, we provide a formulation of MHD in terms of the action
principle.  An action formulation has an advantage of potentially
giving a clear interpretation of conserved quantities as the
consequence of symmetries.  In particular, in our formulation the
conservation of the 2-form current is ensured by a an abelian global
symmetry, in which the parameter of a symmetry transformation is a
1-form, rather than a scalar. This symmetry turns out to be the main
guiding principle to write the most general equations compatible with
the standard assumptions of MHD, i.e., quasi-neutrality of the plasma
and Debye screening of the electric field.  This new symmetry suggests
to include a ``dual photon'' to describe the dynamics of the
electromagnetic part of the fluid. This is natural as the dynamics of
the electromagnetic field is dictated by the conservation equation of
the dual field-strength (\ref{cons0}). The other degrees of freedom
are (the relativistic) Lagrangian coordinates of the fluid.

Previous action formulations of MHD have been implemented mainly by
generalizing the approach of Clebsch potentials
\cite{Bekenstein:2000sf,Webb:2017fgj,Kumar:2017sjg}. Other
descriptions used Lagrangian coordinates \cite{morrison1,kawazura},
although there are no dynamical degrees of freedom associated to the
electromagnetic field.  The advantage of introducing the dual photon
is the presence of an enhanced set of symmetries, which allows one to
systematically find various conserved helicities, providing a more
transparent geometric structure. Another advantage of our formulation
is that it conformes nicely with the philosophy of effective field
theory, in which one writes down systematically all possible terms of
the Lagrangian compatible with the required symmetries, following a
power counting given by derivative expansion. The symmetries guarantee
that each of these terms is consistent with the basic physical
assumptions of MHD. It is worth mentioning that, since this approach
is based on a standard action principle, it will not capture
dissipative contributions, such as conductivity and
viscosities.\footnote{In order to capture dissipation, one needs an
  enhanced variational principle where degrees of freedom are
  doubled~\cite{Crossley:2015tka} (see also Refs.~\cite{Haehl:2015uoc}
  and \cite{Jensen:2017kzi}).}

The paper is organized as follows. We begin Sec.~\ref{sec:act} by
introducing the degrees of freedom and the symmetries, and then
present the general action of ideal MHD and show that the
stress-energy tensor and the dual field-strength, as well as as
possible additional conserved currents, acquire the form expected from
MHD. In Sec.~\ref{sec:ffe} we show how a variational formulation of
force-free electrodynamics (FFE) can be obtained by enhancing one of
the symmetries. We also show that, intriguingly, the equations of FFE
can be recovered as a limit of axion electrodynamics. In
Sec.~\ref{sec:noet} we describe the Noether currents associated with
the symmetries, and show that particular subgroups of such symmetries
give rise to conserved helicities. In Sec.~\ref{sec:high} we discuss
higher-derivative corrections. We state our conclusions in
Sec.~\ref{sec:concl}.

\section{Action principle}\label{sec:act}

We first describe the building blocks for our action formulation: the
degrees of freedom and the symmetries.  We then use these ingredients
to write down the most general action at leading order in derivatives,
which will correspond to ideal MHD, and relate it to thermodynamics.

\subsection{Kinematics}\label{sec:kind}

In the 2-form formulation of Ref.~\cite{Grozdanov:2016tdf}, MHD is
stated in terms of the conservation equations
\be\label{cons}
  \nabla_\mu T^{\mu\nu}=  \frac 12H^{\nu\alpha\beta}J_{\alpha\beta}
  ,\qquad \nabla_\mu J^{\mu\nu}=0\ ,
\ee
where $T^{\mu\nu}$ is the stress-energy tensor and $J^{\mu\nu}$ is a
2-form current which %, for weak electromagnetic fields ({\bf only for
%  weak EM fields?}),
can be identified with the dual field strength
$\tilde F^{\mu\nu}$ defined in (\ref{cons0}). In the above we also
included
\be
  H_{\mu\nu\rho}= \p_\mu b_{\nu\rho}+\p_\nu b_{\rho\mu}+\p_\rho b_{\mu\nu}\ ,
\ee
where $b_{\mu\nu}$ is the background source coupled to
$J^{\mu\nu}$. The field strength $H_{\mu\nu\rho}$ can be thought of as
the dual of an external current which probes the electromagnetic
field, i.e.
\be J^\sigma_{\text{ext}}=- \frac 16\vep^{\sigma\alpha\beta\gamma}
  H_{\alpha\beta\gamma}\ ,\ee
so that
\be
   \frac 12 H^{\nu\alpha\beta}J_{\alpha\beta}= J_{\mu,\text{ext}} F^{\nu\mu}\ ,
\ee
i.e., the right-hand side of the first equation in (\ref{cons}) is the
Lorentz force generated by $J_{\text{ext}}^\mu$.

To find an action $S$ whose equations of motion are the conservation
equations (\ref{cons}), the first step is to require one can couple
the theory to a background metric $g_{\mu\nu}$ and 2-form $b_{\mu\nu}$,
and that
the stress-energy tensor and the 2-form current are obtained by
varying the action with respect to the respective sources, i.e.,
\be \begin{split}\label{curr}
T^{\mu\nu}&=\frac 2{\sqrt{-g}}\frac{\delta S}{\delta g_{\mu\nu}}\\
J^{\mu\nu}&=\frac 2{\sqrt{-g}}\frac{\delta S}{\delta b_{\mu\nu}}\ .\end{split}\ee
The conservation of these currents is then a consequence of the invariance
of the action with respect to diffeomorphism and 1-form $U(1)$:
\be\label{spsym}\begin{split} g_{\mu\nu}(x)&\to \p_\mu y^\alpha \p_\nu y^\beta g_{\alpha\beta}(y(x))\\
  b_{\mu\nu}(x)&\to \p_\mu y^\alpha \p_\nu y^\beta b_{\alpha\beta}(y(x))-2\p_{[\mu}\chi_{\nu]}\ ,\end{split}\ee
where $y^\mu(x)$ and $\chi_\mu(x)$ denote coordinate and one-form
$U(1)$ transformations, respectively.

In analogy with the effective theory formulation of hydrodynamics \cite{Crossley:2015tka,Haehl:2015uoc,Jensen:2017kzi,Dubovsky:2005xd,
Geracie:2014iva,Dubovsky:2011sj,Haehl:2013kra,Glorioso:2018wxw},
we propose that the degrees of freedom
responsible for the conservation equations (\ref{cons}) are the
parameters of such transformations, or more precisely, undergo shifts
under these transformations.
To parameterize diffeomorphisms,
we introduce the fields $\sigma^a(x)$, which we regard as mappings
from the physical spacetime, with coordinates $x^\mu$, to an internal
spacetime, with coordinates $\sigma^a$. To parameterize 1-form $U(1)$
transformations, we introduce the 1-form $\varphi_a(x)$.\footnote{Note
  that we chose to take the indices of $\varphi_a$ to live in the
  $\sigma^a$ spacetime: as we will see later, this choice makes more
  transparent the tensor structure of various equations. Defining the
  1-form $\varphi_\mu$ in physical spacetime leads to an equivalent
  set of equations of motion after the identification
    $\varphi_\mu=\partial_\mu\sigma^a\varphi_a$.} We shall see later
that $\varphi_a$ can be interpreted as a ``dual photon.'' We emphasize
that $\sigma^a,\varphi_a$ are \emph{not} Goldstone modes, as they are not
associated with the spontaneous symmetry breaking of a symmetry
defined in the microscopic system.\footnote{See \cite{Armas:2018atq} for an alternative perspective of this point.}
%In our framework, they arise as
%low-energy modes responsible for the conservation of the currents
%({\bf What does it mean?}).
%\cite{Armas:2018atq}

To ensure that Eqs.~(\ref{cons}) follow from the variation of the
  action with respect to $\sigma^a$ and $\varphi_a$, we demand the
action to depend on the background sources $g_{\mu\nu},b_{\mu\nu}$ and
the degrees of freedom $\sigma^a,\varphi_a$ through the following
combinations:
\be\begin{split} \label{pback}h_{ab}&=K^\mu_{\ a}K^\nu_{\ b} g_{\mu\nu}\\
B_{ab}&=K^\mu_{\ a}K^\nu_{\ b}b_{\mu\nu}+2K^\mu_{\ [a}\p_{\mu}\varphi_{b]}\\
&=K^\mu_{\ a}K^\nu_{\ b}b_{\mu\nu}+2\p_{[a}\varphi_{b]}\end{split}\ee
where $K^\mu_{\ a}=\frac{\p x^\mu}{\p\sigma^a}$.\footnote{One can
    arbitrarily choose either $x^\mu$ or $\sigma^a$ to be the
    dynamical degrees of freedom of the theory responsible for
    energy-momentum conservation.} Requiring
\begin{equation}\label{ShB}
  S=S[h_{ab},B_{ab}]
\end{equation}
ensures that the equations of motion for $\sigma^a$ and $\varphi_a$
imply precisely Eqs.~(\ref{cons}). The combinations (\ref{pback}) also
guarantee that the action is invariant under diffeomorphisms and
1-form $U(1)$ transformations, given by Eqs.~(\ref{spsym}) together
with
\be\begin{split}
\label{diff}\sigma^a(x)&\to \sigma^a(y(x))\\
\varphi_a(x)&\to \varphi_a(y(x))+K^\mu_{\ a}\chi_\mu(x)\ .
\end{split}\ee

Now that we ensured conservation of stress-energy and flux, we come to
the second fundamental ingredient of MHD: The explicit expressions of
$T^{\mu\nu}$ and $J^{\mu\nu}$ should depend only on the fluid velocity
$u^\mu$, the temperature $T$, and the chemical potential for the
1-form charge $\mu h_\mu$.  Following Ref.~\cite{Grozdanov:2016tdf} we
have factorized the 1-form chemical potential into a scalar $\mu$ and
a 1-form $h_\mu$ such that
\be \label{hh} u^\mu h_\mu =0,\quad h^\mu h_\mu =1\ .\ee

The most general action $S$ of the form~(\ref{ShB}) does not satisfy
the above requirement, and the conserved currents, in general, have a
rather generic dependence on $\sigma^a$ and $\varphi_a$.  We now show
that the requirement can be satisfied by imposing additional
symmetries, which, in a certain sense, will truly define the geometric
nature of our MHD theory. In order to introduce these symmetries, we
need to develop intuition on the the physical meaning of the degrees
of freedom $\sigma^a$ and $\varphi_a$ and their relationship with the
hydrodynamic variables $u^\mu$, $T$, and $\mu h_\mu$.

The fields $\sigma^a$ are the relativistic Lagrangian fluid
coordinates.  Heuristically, a fixed value of the spatial coordinates
$\sigma^i$ identifies a given fluid element, whose trajectory in
spacetime is described by the function $x^\mu(\sigma^0,\sigma^i)$ as
$\sigma^0$ varies, and $\sigma^0$ represents the internal ``clock'' of
the fluid element. It is then natural to define the velocity of the
fluid as the normalized tangent vector to such trajectory:
\be\label{vel} u^\mu = \frac 1{b} \frac{\p x^\mu}{\p \sigma^0} \,,
 \qquad b = \sqrt{-h_{00}}\, .
\ee

Consider now a system in a homogeneous configuration at temperature $T_0$. Placing the system in a slowly-varying curved background induces a redshift in the local temperature, so that the latter is $T(x)=T_0/\sqrt{-g_{00}}$. This invites to take the following relation between the local temperature and $\sigma^a$: $T(x)=T_0/\sqrt{-h_{00}}$.  Furthermore, a global rescaling of time $x^0\to \gamma x^0$ induces a rescaling of the asymptotic temperature $T_0\to T_0/\gamma$. We can use this to fix $T_0=1$ and write
\be\label{temp} T(x)=\frac{1}{b}\, .\ee

Finally, $\varphi_a$ can be thought of as a ``1-form phase'' associated to each fluid element, which we want to relate to the 1-form chemical potential. The corresponding charges are the magnetic flux across 2-dimensional surfaces at constant time. In the $\sigma^a$-coordinate system, where $\sigma^0$ singles out the time direction, the charges are
\be\label{Qi} Q^i=\int_{\Sigma_{(i)}} d^2 x\sqrt{-g} J^{0i}\ ,\ee
where $\Sigma_{(i)}$ is the plane at constant values of $\sigma^0$ and $\sigma^i$, for $i=1,2,3$. The thermodynamic partition function is then
\be \label{clf} Z=\Tr(e^{-\frac 1T(H-\mu_i Q^i)})\ ,\ee
where $H$ is the vacuum Hamiltonian of the system, and $\mu_i$ is the chemical potential conjugated to $Q^i$. Comparing Eqs.~(\ref{Qi}), (\ref{clf}), and (\ref{curr}), we are then led to identify $B_{0i}$ with the 1-form chemical potential $\mu_i$ in the $\sigma^a$-coordinates. Pulling back to $x^\mu$ coordinates, this becomes
\be\label{chem1} \mu h_\mu =\frac 1{b} \frac{\p \sigma^i}{\p x^\mu} B_{0i}\ ,\ee
where the prefactor of $\frac 1b$ comes from that $B_{0i}$ behaves like a scalar density in the time direction. We will develop more details on this object in the next section.

We now come to the symmetries. As discussed above, the spatial coordinates $\sigma^i$ can be thought of
%as describing individual fluid elements.
as giving a system of coordinates embedded in the fluid.
However, there is a redundancy in this identification. Given a fluid configuration, one can
%imagine displacing the fluid elements
arbitrarily change the system of coordinates used to parametrize fluid elements
without changing physical quantities, i.e. temperature, velocity and one-form $U(1)$ chemical potential. This is in contrast to, e.g., solids, where
%any non-rigid displacement of lattice sites has a nonzero energy cost.
there is a single-out coordinate systems tied to the lattice.
The redundancy characteristic of fluids is captured by the spatial diffeomorphisms
\be\label{42} \sigma^i\to f^i(\sigma^{j})\ ,\ee
where $f^i$ is an arbitrary function of $\sigma^j$. We then require
that the action should be invariant under (\ref{42}). In other words,
one has the freedom of relabeling fluid elements at a constant time
slice $\sigma^0=\textrm{const}$.  Analogously, the choice of initial
value of the internal time $\sigma^0$ should be arbitrary for each
fluid element. This leads to requiring invariance under time shifts:
\be \label{41}\sigma^0\to \sigma^{0}+f(\sigma^{i})\ .\ee
where $f$ is an arbitrary function of $\sigma^i$. Lastly, one should
also have the freedom of arbitrary choice of the one-form $\varphi_a$
at the initial time. This leads to demanding invariance with respect
to time-independent shifts of the one-form phase
\be \label{43}\varphi_a\to\varphi_a+\lambda_a(\sigma^i)\ ,\ee
where $\psi_a$ is a time-independent one-form with arbitrary dependence on $\sigma^i$.

Finally, given that the action depends on $\varphi_a$ through the second line of Eq.~(\ref{pback}), one has the additional symmetry
\be \label{44} \varphi_a\to \varphi_a+ \p_a \alpha(\sigma)\ ,\ee
where $\alpha$ is an arbitrary scalar function of $\sigma^a$. This symmetry is time-dependent, contrary to the previous ones. We shall see later that it will lead to a generalization of the constraint on the magnetic field $\p_i B^i=0$.

Note that $T,u^\mu,\mu$ and $h_\alpha$, as defined in
Eqs.~(\ref{vel}), (\ref{temp}), and (\ref{chem1}) are invariant under
(\ref{42})--(\ref{43}). In fact, as we will show, the stress-energy
tensor and the 2-form current coming from an action invariant under
(\ref{42})--(\ref{43}), depend only on $T$, $u^\mu$, and
$\mu^\alpha$. This will be a crucial property to verify the
consistency of our formulation.

To make the symmetries (\ref{42})--(\ref{43}) manifest, it is
convenient to parametrize the fields (\ref{pback}) in terms of new
variables $b$, $v_i$, $a_{ij}$, $m_i$, and $m_{ij}$
\be\begin{split}\label{48}
h_{ab}d\sigma^a d\sigma^b=&- b^2(d\sigma^0-v_id\sigma^i)^2
+a_{ij}d\sigma^id\sigma^j\\
B_{ab} d\sigma^a\wedge d\sigma^b =& 2 b m_i(d\sigma^0-v_j d\sigma^j)\wedge d\sigma^i\\
&+m_{ij}d\sigma^i\wedge d\sigma^j\ .
\end{split}\ee
%\begin{gather}\label{48} T=\frac 1{\sqrt{-h_{00}}},\quad v_i=Th_{i0},\quad a^{ij}=h^{ij},\\ \label{49}
%m_i=T B_{0i},\quad m_{ij}=\lambda_i^\mu \lambda_j^\nu B_{\mu\nu},\end{gather}
The objects in (\ref{48}) transform covariantly under
(\ref{42}):
\begin{align}\label{trab}
  & b \to b, \\
  & m_i\to \p_i f^j m_j\,, \quad v_i \to \partial_i f^j v_j\\
  & a_{ij} \to \partial_i f^k \partial_j f^l a_{kl},\quad
   m_{ij} \to \partial_i f^k \partial_j f^l m_{kl}
\end{align}
while under time shifts~(\ref{43}), only $v_i$ transforms nontrivially
\be\label{214} v_i\to v_i-\p_i f,
\ee
and only $m_{ij}$ transforms nontrivially under~(\ref{44}):
\be\label{tram} m_{ij}\to m_{ij}+2\p_{[i}\lambda_{j]}.\ee

From the discussion around (\ref{chem1}) we also see that $m_i$ is the one-form chemical potential in $\sigma^a$-coordinates. In terms of $m_i$, eq. (\ref{chem1}) reads
\be \label{chem}h_\mu =\frac 1\mu \frac{\p \sigma^i}{\p x^\mu}m_i,\qquad \mu=\sqrt{m_i m_j a^{ij}}\ .\ee
Note that the second equation in (\ref{hh}) is automatically guaranteed, indeed
\be u^\mu h_\mu = \frac 1 b \frac{\p x^\mu }{\p \sigma^0} \mu\frac{\p \sigma^i}{\p x^\mu} m_i=\frac \mu b \delta^i_0 m_i =0\ .\ee

\subsection{Ideal MHD}
%In this section we shall extract the thermodynamics coming from the leading order action (\ref{actmhd}). This, in turn, will enable us to relate the variables $\sigma^a,\varphi_a$ to the standard plasma degrees of freedom.
In this Section we shall study the most general action at leading derivative order which is compatible with the symmetry requirements discussed above.
The leading derivative order contains no derivatives at all, meaning that we construct the action based on the fields in the decomposition (\ref{48}). The most general action at this order turns out to depend only on $b$ and $m_i$. Using (\ref{temp}) and (\ref{chem}), it is given by
\be\label{actmhd} S=\int d^4 x\sqrt{-g}F(T,\mu)\ ,\ee
where $F$ is an arbitrary function of $T$ and $\mu$. Varying (\ref{actmhd}) with respect to the background sources as in (\ref{curr}) we obtain stress-energy tensor and two-form current:\footnote{See Appendix \ref{app:for} for the variations of various fields with respect to background sources.}
\be\begin{split}\label{413} T^{\mu\nu}&=\frac 2{\sqrt{-g}}\frac{\delta S}{\delta g_{\mu\nu}}=p g^{\mu\nu}+(\varepsilon+p)u^\mu u^\nu-\mu\rho h^\mu h^\nu\\
J^{\mu\nu}&=\frac 2{\sqrt{-g}}\frac{\delta S}{\delta b_{\mu\nu}}=2\rho u^{[\mu}h^{\nu]},\end{split}\ee
where $p$ and $\vep$ are pressure and energy density, and $\rho$ denotes a scalar associated to the 2-form density. They are related to $F$ through
\be\label{414} p=F,\quad \rho=\p_\mu F,\quad \varepsilon =T\p_T F+\mu\p_\mu F-F \ .\ee
If we identify
\be s=\p_T F\ ,\ee
where $s$ is the entropy density, we find the thermodynamic relation
\be p+\varepsilon=sT+\mu \rho\ ,\ee
which was first obtained in \cite{Grozdanov:2016tdf}.

\subsection{Connection to the traditional formulation of MHD}
In the usual formulation of relativistic ideal MHD \cite{lich}, the stress tensor is the sum of fluid and Maxwell stress tensors:
\be\label{totT} T^{\mu\nu}=T^{\mu\nu}_{\text{fluid}}+T^{\mu\nu}_{\text{maxwell}}\ ,\ee
where
\be\begin{split} T^{\mu\nu}_{\text{fluid}}&=p_0(T) g^{\mu\nu}+(\vep_0(T)+p_0(T)) u^\mu u^\nu\\
T^{\mu\nu}_{\text{maxwell}}&=F^\mu_{\ \alpha}F^{\nu\alpha}-\frac 14 F^2 g^{\mu\nu}\ ,\end{split}\ee
and where one imposes the vanishing of the electric field in the fluid rest frame $F^{\mu\nu}u_\nu=0$, which allows to rewrite
\be T^{\mu\nu}_{\text{maxwell}}=
B^2 u^\mu u^\nu + \frac 12 B^2 g^{\mu\nu}-B^\mu B^\nu\ ,\ee
where $B^\mu=\frac 12 \vep^{\mu\nu\rho\sigma}u_\nu F_{\rho\sigma}$ is the magnetic field. In this case $T^{\mu\nu}_{\text{fluid}}$ is completely disentangled from the electromagnetic part, as the thermodynamic parameters $p_0(T),\vep_0(T)$ are only functions of the temperature. Additionally,
\be \label{dfs}\tilde F^{\mu\nu}=2 u^{[\mu} B^{\nu]}\ ,\ee
and the complete set of equations of motion is given by
\be \nabla_\mu T^{\mu\nu}=0,\quad \nabla_\mu \tilde F^{\mu\nu}=0\ .\ee
We then see that the total stress tensor (\ref{totT}) and the dual field strength (\ref{dfs}) coincide with $T^{\mu\nu}$ and $J^{\mu\nu}$ given in (\ref{413}),(\ref{414}), respectively, upon identifying
\be\label{pp0} p=p_0+\frac 12 B^2%,\quad \vep=\vep_0+\frac 12 B^2,\quad \mu=\sqrt{B^2},\quad \rho=\sqrt{B^2}
\ .\ee
The traditional formulation of MHD can indeed be seen as a particular limit of (\ref{413}), where the dynamics of the electromagnetic field is dominated by the Maxwell Lagrangian. For sufficiently large electromagnetic fields one expects thermal and quantum fluctuations to become important, thus generating a nontrivial dependence on $B^2$ in the pressure, which is captured by the general expressions in (\ref{413}).

\subsection{MHD with number conservation}
In many situations one may need to include an additional conserved charge in MHD. The conservation equation for the corresponding vector current is then
\be \label{con3}\nabla_\mu J^\mu = 0\ .\ee
We emphasize that $J^\mu$ is not the current associated to the electromagnetic $U(1)$. Depending on the context, $J^\mu$ could be the baryon number current, or simply the particle number current, e.g. when dealing with a non-relativistic system, where the particle number is conserved. Such $J^\mu$ can also be the axial current, which is relevant when the system is coupled to chiral matter.\footnote{See Ref.~\cite{Akamatsu:2013pjd} for a discussion on chiral MHD and the dramatic consequences due to quantum anomalies.} We couple the theory to a background gauge field $C_\mu$ so that $J^\mu$ is obtained by the variation
\be J^\mu=\frac1{\sqrt{-g}}\frac{\delta S}{\delta C_\mu}\ .\ee
Note that in the presence of $J^\mu$, the first eq. in (\ref{cons}) is modified to
\be\label{cons2} \nabla_\mu T^{\mu\nu}=\frac 12 H^{\nu\alpha\beta}J_{\alpha\beta}+D^{\nu\alpha}J_\alpha\ ,\ee
where $D_{\mu\nu}=\p_\mu C_\nu-\p_\nu C_\mu$, and the conservation of the 2-form is unaffected.

The conservation equation (\ref{con3}) is associated to invariance of $S$ under the background gauge transformation
\be \label{u1g}C_\mu\to C_\mu + \p_\mu \lambda\ ,\ee
where $\lambda(x)$ is a function. In parallel with sec. \ref{sec:kind}, we then introduce a scalar field $\varphi(x)$ which parameterizes such transformations. To ensure that eq. (\ref{con3}) follows from the varying the action with respect to $\varphi$, we require that the action depends on $C_\mu$ and $\varphi$ through the combination
\be \label{Ba}B_a= K_a^\mu(C_\mu+\p_\mu\varphi)\ .\ee
This combination ensures that the action will be invariant under the symmetries discussed before, as well as (\ref{u1g}), provided we simultaneously transform
\be \varphi\to \varphi-\lambda\ .\ee
Finally, we demand the action to be invariant under the transformation \be\label{cs}
\varphi\to\varphi+\chi(\sigma^i)\ee
(keeping $C_\mu$ fixed), where $\chi(\sigma^i)$ is an arbitrary function of the spatial fluid coordinates. This corresponds to requiring the freedom of relabeling the $U(1)$ phase of each fluid element at a given time slice. The same symmetry was required in the formulation of charged fluids in \cite{Dubovsky:2011sj}, where the transformation (\ref{cs}) was dubbed ``chemical shift''.

It will be convenient to decompose $B_a$ as
\be\label{mn} B_a d\sigma^a=b\mu_N(d\sigma^0-v_id\sigma^i)+b_id\sigma^i\ ,\ee
where we shall interpret $\mu_N$ as the chemical potential associated to the $U(1)$ charge. This can be motivated by a similar argument as that around eq. (\ref{chem1}). The variables $\mu_N$ and $b_i$ transform covariantly under (\ref{42})-(\ref{44}), while under (\ref{cs}),
\be \mu_N\to \mu_N,\qquad b_i\to b_i+\p_i\chi\ .\ee
At ideal level, the most general action is then
\be S=\int d^4 x \sqrt{-g}\, F(T,\mu,\mu_N)\ ,\ee
where $F$ is an arbitrary function. Varying $S$ with respect to $C_\mu$ gives the current
\be \label{JJ}J^\mu = n u^\mu\ ,\ee
where $n=\p_{\mu_N} F$ is the number density. The other constitutive relations (\ref{413}) are unmodified, as well as (\ref{414}), except that now
\be \vep=T\p_T F+\mu\p_\mu F+\mu_N\p_{\mu_N}F-F\ ,\ee
leading to the thermodynamic relation
\be p+\vep=sT+\mu\rho+\mu_N n\ .\ee

\section{Force-free electrodynamics}\label{sec:ffe}

An important limit of MHD if force-free electrodynamics. In this limit the contribution of electromagnetic fields to the dynamics is much more relevant than the contribution of matter, the latter essentially decouples from the equations but the nonlinear structure of MHD remains. This is a very useful approximation in astrophysics to study e.g. the solar corona or other types of astrophysical plasma, such as black hole and neutron star atmospheres \cite{fleish,Gralla:2014yja}.

In FFE, the stress-energy tensor is approximated with that coming from Maxwell Lagrangian $-\frac 14 F^{\mu\nu}F_{\mu\nu}$, which implies that
\be F^{\mu\nu}J_\mu=0\ ,\ee
where $J^\mu$ is the electromagnetic current produced by the plasma. Using Maxwell's equations, we can then write FFE in the following form
\be \label{ffe0}\nabla_\mu \tilde F^{\mu\nu}=0,\qquad \nabla_\mu F^{\mu\nu}F_{\nu\rho}=0\ ,\ee
where we are left with are left with a non-linear system which determines the evolution of $F_{\mu\nu}$.

Below we shall see that an action for FFE can be obtained by enhancing the symmetry requirements discussed in the previous Section. We will also show that, intriguingly, eqs. (\ref{ffe0}) can be obtained from a limit of axion electrodynamics.

\subsection{FFE as symmetry-enhanced MHD}

We now require, instead of (\ref{41}), that the action be invariant under the larger group of transformations
\be\label{41a} \sigma^0\to f(\sigma^0,\sigma^i)\ ,\ee
where $f$ is an arbitrary function. Clearly, this transformation implies that $\sigma^0$ decouples from the action. Under (\ref{41a}), the temperature (\ref{temp}) transforms as
\be T\to T\left(\frac{\p f}{\p\sigma^0}\right)^{-1}\ ,\ee
and thus, the most general action at leading derivative order is
\be\label{ffel} S=\int\! d^4 x \, \sqrt{-g}\,F(\mu)\ .\ee
The stress-energy tensor and two-form current obtained from this action are the same as those in Eqs.~(\ref{413}) and (\ref{414}), except that now $\p_TF=0$. The interesting observation is that such system can be shown to be equivalent to a generalization FFE.

Identifying $J^{\mu\nu}=\frac 12 \varepsilon^{\mu\nu\alpha\beta} F_{\alpha\eta}$, the Maxwell energy-momentum tensor can be written as
\be \label{str1}T^{\mu\nu}=-\frac 14 J^2 g^{\mu\nu}+J^{\mu\alpha} J^{\nu}_{\ \alpha}\ ,\ee
which has  the form of (\ref{413}) upon using (\ref{414}), with  $F(\mu)=\frac 12 \mu^2$. The action (\ref{ffel}) can then be viewed as the generalization of FFE to non-linear electrodynamics.  We also note that the structure of the above constitutive relations has similarities with the zero-temperature limit of \cite{Grozdanov:2016tdf}.

\subsection{FFE as a limit of axion electrodynamics}

In this section we describe a completely different action principle
for FFE, which can be thought of as a limit of axion electrodynamics.
This action is
\be \label{axe}S=-\frac 14 \int\! d^4 x\,\sqrt{-g}\left(
F^{\mu\nu}F_{\mu\nu}+\kappa\theta
\vep^{\mu\nu\alpha\beta}F_{\mu\nu}F_{\alpha\beta}\right)\ ,\ee where
$\kappa$ is a constant, and we regard the axion $\theta$ to be
dynamical.  Note that both the kinetic term and the potential term for
$\theta$ are absent, so $\theta$ is now a Lagrange multiplier.

Varying $A_\mu$ and $\theta$, one finds the equations of motion
\be \label{ffee}\begin{split}\vep^{\mu\nu\alpha\beta}F_{\mu\nu}F_{\alpha\beta}
  &=0\\
\nabla_\mu F^{\mu\nu}&=J^\nu\ ,\end{split}\ee
where the current $J^\mu$ is defined by
\be \label{curf}J^\mu \equiv \kappa \vep^{\mu\nu\alpha\beta}\p_\nu \theta F_{\alpha\beta}\ .\ee
Now, using the identity
\be V_\mu \vep^{\alpha\beta\gamma\delta}F_{\alpha\beta}F_{\gamma\delta} =-4\vep^{\alpha\beta\gamma\delta}F_{\mu\alpha}V_\beta F_{\gamma\delta}\ ,\ee
one finds
\be\begin{split} \label{orth}F_{\rho\nu}J^\nu&=\kappa\vep^{\mu\nu\alpha\beta}\p_\nu\theta F_{\rho\mu}F_{\alpha\beta}\\
&=-\frac 14 \kappa \p_\rho\theta \vep^{\mu\nu\alpha\beta}F_{\mu\nu}F_{\alpha\beta}=0\ .\end{split}\ee
We thus recovered precisely the equations of force-free
electrodynamics! The current in Eq.~(\ref{curf}) may appear to possess a
more specific form than that in FFE, but it turns out that this not
the case. Indeed, from Eq.~(\ref{orth}) we know that $F_{\mu\nu}$, being
antisymmetric, has rank 2, and thus one can write
\be F_{\mu\nu}=V_{[\mu}W_{\nu]},\quad J^\mu V_\mu=J^\mu W_\mu=0\ ,\ee
where $V_\mu,W_\mu$ are two vector fields. Hence, there exist a vector field $S_\mu$ such that
\be J^\mu=\vep ^{\mu\alpha\beta\gamma}V_\alpha W_\beta S_\gamma=\vep ^{\mu\alpha\beta\gamma}F_{\alpha\beta} S_\gamma\ ,\ee
and, due to the Bianchi identity,
\be \nabla_\mu J^\mu=\vep ^{\mu\alpha\beta\gamma}F_{\alpha\beta}\nabla_\mu S_\gamma=0\ ,\ee
which in turn implies $\nabla_{[\mu}S_{\nu]}=0$. We then must have
$S_\mu = \p_\mu \alpha$, for some function $\alpha$. This then shows
that Eq.~(\ref{curf}) describes the most general current in FFE, and we
conclude that the equations from the axion electrodynamics action
(\ref{axe}) are completely equivalent to FFE.

The action~(\ref{axe}) should be considered as a formal rewriting of
FFE, and $\theta$ as an unphysical Legendre multiplier.  One may ask,
however, if there is a regime in which axion electrodynamics, with an
axion field of interest for particle physics, reduces to FFE.  We now
show that the neglection of the kinetic term and the mass term in the
Lagrangian~(\ref{axe}) can be justified only at very large values of
the magnetic field.  To see that, we add these terms into the
Lagrangian
\begin{equation}
  -\frac {f_a^2}2 [(\partial_\mu\theta)^2 + m^2\theta^2]
\end{equation}
where $f_a$ and $m_a$ are the decay constant and mass of the axion,
respectively, and requires that the effect of these new terms on the
equation of motion is negligible.  Consider a mostly magnetic solution
where $E\ll B$.  From an equation of motion
\begin{equation}
  \partial_i E^i = \kappa B^i \partial_i\theta
\end{equation}
it follows that $\theta\sim E/(\kappa B)$.  The new terms in the
action can then be neglected when
\begin{equation}
  f_a^2 \frac{\theta}{L^2}\,,\  f_a^2 m_a^2 \theta \ll \kappa EB
\end{equation}
where $L$ is the length scale over which the fields vary.  This
condition can be rewritten as
\begin{equation}
  B \gg \frac{f_a m_a}\kappa\,,\  \frac{f_a}{\kappa L}
\end{equation}
For QCD axions $f_ama_a\sim \Lambda_{\rm QCD}^2$ and
$\kappa\sim\alpha$, and one needs magnetic fields stronger than
$10^{20}$ G.  We leave open the possibility that FFE can be realized
in some regime of the axion electrodynamics relevant for certain
time-reversal breaking topological insulators, where $\theta$ plays
the role of a dynamical magnetization
\cite{Li:2009tca,Ooguri:2011aa,axion3}.

\begin{comment}
It is interesting to compare (\ref{axe}) to the effective action for the magnetization considered in \cite{Li:2009tca} in which, in our conventions, $\kappa=\frac \alpha\pi$, where $\alpha$ is the fine-structure constant, and to $S$ in (\ref{axe}) we add the kinetic term for the axion
\be S_\theta=g^2 J\int d^4 \left[(\p_0\delta\theta)^2- v^2 (\p_i \delta\theta)^2-m^2\delta\theta^2 \right]\ ,\ee
where the axion decay constant $g^2 J$, as well as the mass $m$ and the velocity $v$ are obtained from the microscopic model, and $\delta\theta=\theta-\theta_0$, where $\theta_0$ is some equilibrium value. The equation of motion for the axion is now modified to
\be\label{axion4} -\frac 14 \frac \alpha\pi \vep^{\mu\nu\alpha\beta}F_{\mu\nu}F_{\alpha\beta}=g^2 J(\p_0^2-v^2\p_i^2+m^2)\delta\theta\ . \ee
In order for (\ref{axion4}) to be appreciably close to (\ref{ffee}) we then must have
\be \alpha E_i B_i\gg g^2 J m^2\ ,\ee
where we assumed that the frequency and momentum of $\delta\theta$ are smaller than the mass $m$. Plugging in typical values \cite{Li:2009tca},
\be \frac{\alpha  E_i B_i}{g^2 J m^2}\sim 10^{19}\ , \ee
thus giving compatibility with (\ref{ffee}).
\end{comment}

\section{Noether charges}\label{sec:noet}

Equations (\ref{42})--(\ref{43}) and (\ref{cs}) constitute infinitely many distinct symmetries, which will give infinitely many conserved currents.  These can be conveniently written in the $\sigma^a$-coordinates:
\begin{eqnarray}
\label{fJ}J^a_f&=&\sqrt a s f\delta^a_0\\
\label{etaJ}J^a_{\eta^i}&=&\sqrt a\left(s v_i+n b_i+\frac{\rho}\mu m^j m_{ij}\right)\eta^i\delta^a_0\\
\label{lJ}J^a_{\lambda_i}&=&\frac{\sqrt a\rho}\mu m^i\lambda_i\delta^a_0\\
\label{nJ} J^a_\chi&=&\sqrt a n \chi\delta^a_0\ ,
\end{eqnarray}
where $f,\lambda_i$ and $\chi$ are the functions introduced in (\ref{41}), (\ref{43}) and (\ref{cs}), respectively, and $\eta^i$ is an infinitesimal transformation defined through $f^i=\sigma^i+\eta^i$, where $f^i$ was introduced in (\ref{42}). Conservation of the currents is then $\p_a J^a=0$. The conservation of the first two currents is entropy and momentum conservation. Conservation of the third current is a generalization of the ``frozen-in'' condition for the magnetic field. The last one corresponds to number conservation. The symmetry (\ref{44}) gives the constraint
\be \label{gaugec}\p_i\left(\frac{\sqrt a \rho}\mu m^i \right)=0\ ,\ee
which
%at weak coupling
reduces to the constraint on the magnetic field $\p_i B^i=0$. One can easily verify that the conservation of (\ref{fJ})-(\ref{nJ}) together with (\ref{gaugec}) are equivalent to the equations of motion (\ref{cons}),(\ref{con3}) and (\ref{cons2}), so (\ref{fJ})-(\ref{nJ}) constitute a complete basis of conserved quantities.

The currents (\ref{fJ})-(\ref{nJ}) are all non-local when written in terms of the Eulerian variables $T,\mu,\mu_N, u^\mu$ and $h^\mu$, in the sense that (\ref{fJ})-(\ref{nJ}) have a dependence on the initial values of such variables. For example, (\ref{etaJ}) depends on the degree of freedom $\varphi$ through $b_i=\p_i\varphi+v_i\p_0\varphi$, where we used (\ref{Ba}) and (\ref{mn}), and we switched off the gauge field $C_\mu=0$ for simplicity. Using (\ref{Ba}) and (\ref{mn}), one has $\frac{\p \varphi}{\p\sigma^0}=\mu_N/T$, which allows to express $\varphi$ in terms of Eulerian variables,
\be\label{phimu} \varphi(\sigma^0,\sigma^i)= \int_{-\infty}^{\sigma^0}d\zeta\,\frac{\mu_N(\zeta,\sigma^i)}{T(\zeta,\sigma^i)}\ ,\ee
where we took $\varphi=0$ in the far past. We then see that (\ref{etaJ}) depends non-locally on the usual hydrodynamic variables. As we will see, this non-locality is related to the fact that the currents (\ref{fJ})-(\ref{nJ}) are not invariant under some of the symmetries (\ref{42})-(\ref{44}). The issue of finding conserved helicities and understanding if they can be written locally in Eulerian variables is a topic of interest in MHD \cite{Bekenstein:2000sf,Webb:2017fgj,webb}; helicities are useful to characterize solutions and have wide applications to numerical simulations. The discussion of this subsection aims at giving a systematic derivation of local and non-local helicities using the relation between locality of a given conserved quantity and invariance of the latter with respect to the symmetries (\ref{42})--(\ref{43}) and (\ref{cs}). Below we will consider specific subgroups of our symmetries and find that they recover a generalization of known conserved helicities, and we will see how these helicities become local in specific limits.

Let us start by considering fluid relabelings of the form
\be \label{sdiff}\eta^i=\frac 1{\sqrt a s}\ep^{ijk}\p_j z_k,\quad z_k=z_k(\sigma^i)\ ,\ee
where we included $\sqrt a$ so that $\eta^i$ is a spatial vector (rather than a vector density).\footnote{We define $\ep^{123}=1$.} The independence of $\eta^i$ on $\sigma^0$ is guaranteed as far as the conservation of (\ref{fJ}) is imposed. Note that, up to the prefactor of $\frac 1{\sqrt a s}$ which can be fixed to 1 using (\ref{42}), $\eta^i$ in (\ref{sdiff}) is divergenceless, i.e. it is the infinitesimal generator of volume-preserving diffeomorphisms, or special diffeomorphisms.\footnote{More explicitly, one can fix the fluid relabeling symmetry (\ref{42}) by requiring $\sqrt a s=1$ on a given constant-$\sigma^0$ slice. Thanks to (\ref{fJ}), this condition is then preserved along all the flow, so that one recovers the well-known special diffeomorphism invariance of ideal fluids.} The associated charge is obtained by integrating over a $\sigma^0=c$ slice, where $c$ is a constant,
\be \label{H1b0}H_{\text{sdiff}}=\int d^3\sigma \left( v_i+\frac ns b_i+\frac{\rho}{s\mu} m^l m_{il}\right)\ep^{ijk}\p_j z_k\ .\ee
Under (\ref{trab})-(\ref{tram}), the above quantity transforms as
\be\begin{split} H_{\text{sdiff}}\to &\int d^3\sigma \left(v_i- \p_i f\right.\\
&\left.+\frac ns (b_i+\p_i \chi)+\frac{\rho}{s\mu} m^l (m_{il}+2\p_{[i}\lambda_{l]})\right)\ep^{ijk}\p_j z_k\ .\end{split}\ee
Invariance then imposes the conditions:
\be\label{local0}\begin{gathered} \p_i\left(\frac ns\right)\ep^{ijk}\p_j z_k=0,\quad
\p_i\left(\frac \rho{s\mu} m^{[l}\ep^{i]jk}\p_j z_k\right)=0 \ .\end{gathered}\ee
When these conditions are met, $H_{\text{sdiff}}$ can be written as a local quantity. To show this, it is convenient to choose a gauge for the symmetry (\ref{41}) in which $\sigma^0=x^0$ on the slice $\sigma^0=c$, upon which $\p_I\sigma^0,\p_i x^0=0$, and
\be \ep^{ijk}=\frac{\ep^{IJK}}J\p_I\sigma^i\p_J \sigma^j\p_K \sigma^k\ ,\ee
where $I,J,K$ denote spatial indices in the $x^\mu$ spacetime, and where $J$ is the determinant of $\frac{\p\sigma^i}{\p x^J}$, and moreover
\be \p_\alpha \sigma^l v_l=\frac 1b u_\alpha+ \p_\alpha\sigma^0\ ,\ee
and, for simplicity, we set all the background sources to zero, i.e. $g_{\mu\nu}=\eta_{\mu\nu}$, $b_{\mu\nu}=0$ and $C_\mu=0$. Pulling-back (\ref{H1b0}) to $x^\mu$-coordinates then gives
\be\label{H1c} H_{\text{sdiff}}^{(\text{isen})}=\int d^3 x\, \ep^{IJK}\p_J\left(\frac {h+\rho\mu} s u_I-\frac{\mu u_L B_L}{s\rho(u^0)^2}B_I\right)z_K\ ,\ee
where $h=Ts+\mu_N n$ is the enthalpy density, and $B_\mu=\p_\mu\sigma^i \frac\rho \mu m_i$ is the magnetic field. One physically relevant case in which condition (\ref{local0}) is satisfied is that of isentropic fluids, for which $n/s=$const. \cite{Gourgoulhon:2006bn}, with no electromagnetic contribution $m^l m_{il}=0$. We leave it as an open question whether there are other physically interesting situations where (\ref{local0}) is satisfied.

Let us now set again $m^l m_{il}=0$. As pointed out above, if the fluid is not isentropic the conserved helicity (\ref{H1b0}) is not a local expression of Eulerian variables. To understand this better let us take again $m^l m_{il}=0$ for simplicity, without assuming isentropicity. The pull-back to physical spacetime of $H_{\text{sdiff}}$ is
\be\begin{split} H_{\text{sdiff}}=&\int d^3 x\, \ep^{IJK}\p_J\left(\frac h s u_I\right)z_K\\
&+\int d^3 x \frac n{s}\p_I\varphi \ep^{IJK}\p_J z_K\ .\end{split}\ee
Clearly, if $n/s$ is not constant the second term will in general contribute, making $H_{\text{sdiff}}$ non-local when expressed in terms of Eulerian variables, as discussed around (\ref{phimu}). See Ref.~\cite{webb} for an alternative discussion of this non-local invariant in the case of vanishing electromagnetic fields $m^l m_{il}=0$. A derivation of Kelvin's theorem for fluids from symmetry was discussed in Ref.~\cite{Dubovsky:2005xd}. The conservation of (\ref{H1b0}) is the extension of Kelvin's circulation theorem to relativistic magnetohydrodynamics, and includes the non-barotropic case.

Now let us consider the following transformations
\be\eta^i(\sigma^j)=\eta_0(\sigma^j) \frac 1{\sqrt a s}\frac{\sqrt a \rho}\mu m^i,\qquad m^i\p_i\eta_0=0\ .\label{228}\ee
These fluid relabelings generate a shift along the magnetic field, where the amount of shift is the same along the same magnetic field line. We can thus view them as  ``magnetic shifts.'' Time-independence of $\eta^i$ is guaranteed after accounting for the conservation of (\ref{fJ}) and (\ref{lJ}), which imply that $\frac 1{\sqrt a s}$ and $\frac{\sqrt a \rho}\mu m^i$ are both time-independent. The conserved charge associated to this symmetry is
\be\label{heta} H_{\text{mshift}}\equiv\int d^{3}\sigma \left( v_i+\frac n s b_i\right) \frac{\sqrt a \rho} \mu m^i \eta_0\ .\ee
Let us assume again that the plasma is isentropic $n/s=$const. The above charge is invariant under all the symmetries (\ref{42})-(\ref{44}). In particular, under (\ref{41}) one finds, using the first eq. in (\ref{214}),
\be H_{\text{mshift}}\to H_{\text{mshift}}-\int d^{3}\sigma\,  \p_i f \frac{\sqrt a \rho} \mu m^i \eta_0\ ,\ee
where the integrand can be easily shown to be a total derivative, thanks to the conservation of (\ref{gaugec}), leading to the vanishing of the integral. The same holds for transformation (\ref{cs}) acting on $b_i$. Since $\eta_0$ is an arbitrary function satisfying the second eq. in (\ref{228}), the conservation of $H_{\text{mshift}}$ is equivalent to the conservation of
\be \int_{\mathcal M} d\ell\,\frac \rho\mu m^i \left( v_i+\frac ns b_i\right)\ ,\ee
where $\mathcal M$ is a line in the $\sigma^i$-space generated by the vector $m^i$, and $d\ell$ is the associated line element. In other words, for each magnetic flux tube we have an associated conserved quantity. Pulling back to $x^\mu$ spacetime, eq. (\ref{heta}) becomes
\be \label{heta1} H_{\text{mshift}}=\int d^3 x\,  h \mathfrak b^I u^I \eta_0\ ,\ee
where $h$ is the enthalpy density defined before, $\mathfrak b^I=J^{0I}$ is the magnetic field in the ``lab frame'', and where we set again to zero background sources, $g_{\mu\nu}=\eta_{\mu\nu}$, $C_\mu=0$, $b_{\mu\nu}=0$.
%At weak coupling, when
Eq.~(\ref{heta1}) is precisely the standard cross-helicity for an isentropic fluid. Relaxing isentropicity, eq. (\ref{heta}) cannot in general be written as a local quantity, and gives the cross-helicity for general MHD. A discussion on non-local cross-helicity for traditional MHD can be found in \cite{webb}.

Lastly, we consider the generalized FFE action (\ref{ffel}). There is an accidental infinitesimal symmetry in this case, given by the following time-dependent fluid relabelings $\eta^i$:
\be \eta^i=\frac{\ep^{ijk}}{\sqrt a \mu \rho}m_j \p_k g,\qquad \p_0 g=0,\quad m^i\p_i g=0\ .\ee
The associated conserved charge is
\be \begin{split}H_{\text{ffe}}&=\int d^3\sigma J^0_{\eta^i}=\int d^3\sigma \frac{\sqrt a\rho}\mu m_{ij}m^i \eta^j\\
&=\frac 12 \int d^3\sigma m_{ij}\ep^{ijk}\p_k g=-\int d^3\sigma \ep^{ijk}\p_k m_{ij} g\ ,\end{split}\ee
where we used the identity
\be \ep^{ijk}M_{il}V_j W_k=\ep^{ijk}W_{[i}V_{l]} M_{jk}\ ,\ee
where $M_{jk}$ is an antisymmetric tensor. Upon using the equations of motion, this quantity is time-independent:
\be \p_0 H_{\text{ffe}}=\int d^3\sigma \p_0 m_{ij}\ep^{ijk}\p_k g=0\ ,\ee
where we used that $\p_0\left(\frac{\sqrt a \rho}{\mu} m^j m_{ij}\right)=\frac{\sqrt a \rho}{\mu} m^j \p_0m_{ij}=0$, and $m^i\p_i g=0$, together with the antisymmetry of $\ep^{ijk}$. To get more physical intuition on this quantity, let us specify to Maxwell force-free electrodynamics. Pulling-back to physical space-time and again fixing $\sigma^0=x^0$ on the hypersurface of integration, we find
\be\begin{split} H_{\text{ffe}}&=-\int d^3 x  g\ep^{IJK} \p_I J_{JK} \\
&=-2\int d^3 xg\p_I E^I=-2\int d^3 x g \rho_q\ ,\end{split}\ee
where $\rho_q$ is the electric charge of the system. In the above steps, we used that $J^{\mu\nu}=\frac 12 \ep^{\mu\nu\rho\sigma} F_{\rho\sigma}$ and Gauss' law $\p_i E^i=\rho_q$. In other words, $H_{\text{ffe}}$ is the electric charge integrated along magnetic flux tubes.

\section{Action at first order}\label{sec:high}
In this section we discuss first-derivative contributions to the MHD action. This entails writing the list of couplings which contain one derivative and are invariant under the symmetries discussed in sec. \ref{sec:kind}. To make the symmetries (\ref{42})-(\ref{44}) more manifest, it will be convenient to use derivatives performed with respect to $\sigma^a$ instead of $x^\mu$. To this aim, we introduce a ``covariant'' derivative in the $\sigma^a$-spacetime. Given a scalar $\varphi(\sigma^a)$, its ordinary $\sigma^i$-derivative is not invariant under (\ref{41}):
\be \p_i\varphi\to \p_i\varphi+\p_i f\p_0\varphi\ .\ee
This has the opposite transformation compared to $v_i$ in (\ref{214}), and thus we are lead to introducing the covariant derivative
\be d_i\varphi\equiv \p_i\varphi+v_i\p_0\varphi\ .\ee
One can verify that $d_i\varphi$ transforms as a vector under (\ref{42}), and does not transform under any of the other symmetries. The time derivative $\p_0\varphi$ does not transform under any symmetry.%\footnote{Consider a spatial vector $\varphi_i$, i.e. $\varphi_i$ transforms as a vector under (\ref{42}), and does not transform under any other symmetry. One can extend the definition of $d_i$ to its action on $\varphi_i$ by demanding that $d_i \varphi_j$ transforms as a tensor under (\ref{42}), and that $d_i$ satisfies the Leibnitz rule. We will not need this object below, and remind the interested reader to \cite{CGL}.}

We can now write down the general MHD action at first derivative order, which is given by:
\be\begin{split}\label{439} S_{(1)}=&\int d^4 x\sqrt{-g}\big( a_1 \vep^{ijk} m_i d_j v_k\\
&+a_2T\vep^{ijk}\p_0 m_{ij} v_k+a_3 \vep^{ijk}d_i m_{jk}\big)\ ,\end{split}\ee
where $\vep^{ijk}=\ep^{ijk}/\sqrt a$. For simplicity, we considered MHD without additional conserved currents, and we included only terms that are linear in the electromagnetic field. We also restricted to terms that satisfy parity and time-reversal invariance. See  Appendix \ref{app:1mhd} for MHD with one additional conserved current. Here, $a_1$ and $a_3$ are arbitrary functions of $T$. The term multiplying $a_2$ has a non-trivial transformation under (\ref{41}). In $\sigma^a$-coordinates:
\be\label{act1}\begin{split}& \delta\int d^4 \sigma a_2\ep^{ijk}\p_0 m_{ij} v_k= -\int d^4 \sigma a_2\vep^{ijk}\p_0 m_{ij} \p_k f\\
&=\int d^4\sigma\left((\p_0 a_2)\vep^{ijk} m_{ij} \p_k f-\p_0(a_2\vep^{ijk} m_{ij} \p_k f)\right).\end{split}\ee
In order for the above to vanish we then require $a_2$ to be a constant, independent of $T$. It is interesting that the action formulation constraints $a_2$ to be a constant. A similar type of restriction was observed before in \cite{Geracie:2014iva}. Comparing with the table below, we see that all the couplings in (\ref{act1}), being linear in the electromagnetic field break charge conjugation, while parity and time-reversal are preserved. The coefficients $a_1$ and $a_3$ survive in the static limit, and are thus of thermodynamic nature.

\begin{center}
\begin{tabular}{|c||c|c|c|c|c|c|}%{ |p{2cm}||p{3cm}|p{2cm}|p{2cm}|p{2cm}|p{2cm}| p{2cm}|}
  \hline
 & $\quad T\quad$ &  $\quad \mu\quad$ & $\quad v_i\quad$ & $\quad a_{ij}\quad$ & $\quad m_i\quad$ & $\quad m_{ij}\quad$ \\
  \hline
$\ T\ $   & $+$&$+$&$-$&$+$&$-$&$+$ \\
$\ P\ $   & $+$&$+$&$-$&$+$&$+$&$-$ \\
$\ C\ $   &  $+$&$+$&$+$&$+$&$-$&$-$\\
 \hline
\end{tabular}
\end{center}

Using the variations given in appendix \ref{app:for}, we find the following first order corrections to the constitutive relations:
\be\label{456}\begin{split}
T^{\mu\nu}_{(1)}=&\delta\vep u^\mu u^\nu+2 u^{(\mu}q^{\nu)}\\
J^{\mu\nu}_{(1)}=&2u^{[\mu}\kappa^{\nu]}+j^{\mu\nu}\ ,
\end{split}\ee

where $q^\mu,\kappa^\mu,j^{\mu\nu}$ are transverse to $u^\mu$, and\footnote{Given that we consider only linear order in the magnetic field, we refrain from further decomposing those tensors into transverse tensors to both $u^\mu$ and $h^\mu$.}
\be\begin{split}\label{const1}
\delta\vep=&-\tfrac 16 (T\p_T a_3-a_3)\vep^{\alpha\beta\gamma\delta} u_\alpha
H_{\beta\gamma\delta}\\
&+\mu (T^2\p_T a_1+2 T\p_T a_3-2 a_3)h_\alpha \Omega^\alpha\\
%q^\mu=&(T\p_T a_1-a_1+2\p_T a_3-2a_3/T)s_1^\mu-T a_1 \mathcal B^\mu\\
%&+2(a_1+a_2+a_3/T) v_1^\mu-(a_2 T+a_3) M^\mu\\
q^\mu=&-(T\p_T a_1-a_1+2\p_T a_3-2\tfrac{a_3}T)s_1^\mu+T a_1 \mathcal B^\mu\\
&-2\left(a_1+(1-\tfrac{sT}{\rho\mu})(a_2+\tfrac {a_3}T)\right) v_1^\mu\\
&-\frac 1{6}(a_2 T+a_3) \vep^{\alpha\beta\gamma\delta}h_\alpha H_{\beta\gamma\delta}h^\mu\\
&+2(a_2 T+a_3)\mathcal B^\alpha h_\alpha h^\mu\\
\kappa^\mu=&a_1T\Omega^\mu\\
j^{\mu\nu}=&2(a_2+\p_T a_3) \vep^{\mu\nu\alpha\beta} u_\alpha \p_\beta T\\
&+2(T a_2+ a_3 ) \vep^{\mu\nu\alpha\beta} u_\alpha \p u_\beta
\ ,\end{split}\ee
%\be\begin{split}
%q^\mu=&(T\p_T a_1-a_1)s_1^\mu+T\p_\mu a_1 s_2^\mu+T\p_{\mu_N}a_1 s_3^\mu\\
%&+2 a_1 v_1^\mu-T a_1 \mathcal B^\mu+2a_2 v_1^\mu+a_2 T M^\mu\\
%\kappa^\mu=&a_1T\Omega^\mu
%\ ,\end{split}\ee
with
\be\begin{gathered}
v_{1}^\mu=\mu\varepsilon^{\mu\alpha\beta\delta}u_\alpha h_\beta(\p_\delta T+T \p u_\delta)\\
\Omega^\mu=\varepsilon^{\mu\alpha\beta\gamma} u_\alpha \p_\beta u_\gamma,\quad
\mathcal B^\mu=\varepsilon^{\mu\alpha\beta\gamma} u_\alpha \p_\beta (\mu h_\gamma)\\
s_1^\mu=\mu\varepsilon^{\mu\alpha\beta\gamma} u_\alpha h_\beta \p_\gamma T%,\quad s_2^\mu=\varepsilon^{\mu\alpha\beta\gamma} u_\alpha B_\beta \p_\gamma \mu
%\\
%M^\mu=\varepsilon^{\mu\alpha\beta\gamma} u_\alpha ( u^\rho H_{\rho\beta\gamma}+2 \p_\beta B_\gamma-2 B_\beta\p u_\gamma)
\ .
\end{gathered}\label{ts1}\ee
The tensor structures in (\ref{ts1}) are all independent after imposing the ideal equations of motion. We rewrite these constitutive relations in the hydrodynamic frame used in \cite{Grozdanov:2016tdf,Hernandez:2017mch}:
\be \label{frame1}\begin{split}
T_{(1)}^{\mu\nu}&=\delta f \tilde \Delta^{\mu\nu}+\delta \tau h^\mu h^\nu+2\ell^{(\mu} h^{\nu)}+t^{\mu\nu}\\
J_{(1)}^{\mu\nu}&= 2m^{[\mu}h^{\nu]}+s^{\mu\nu}\ .
\end{split}\ee
To go from frame (\ref{456}) to frame (\ref{frame1}) we redefine
\be\begin{gathered} u^\mu\to u^\mu+\delta u^\mu,\quad h^\mu+\delta h^\mu\\ T\to T+\delta T,\quad \mu\to \mu+\delta \mu\ ,\end{gathered}\ee
where $\delta u^\mu,\delta h^\mu,\delta T,\delta\mu$ are first order expressions, with $u_\mu \delta u^\mu=h_\mu \delta h^\mu=0$. For a suitable choice of these expressions, we find
\be \label{compar}\begin{split}
\delta f&=\tfrac 16 \left(\tfrac{\p p}{\p \vep}\right)_\rho T^2\p_T\tfrac{a_3}T\vep^{\alpha\beta\gamma\delta}u_\alpha H_{\beta\gamma\delta}
-T\left(\left(\tfrac{\p p}{\p \rho}\right)_{\vep} \tfrac{a_1}\mu\right.\\
&\left.+\left(\tfrac{\p p}{\p \vep}\right)_\rho T\mu\p_T( a_1+2\tfrac{a_3}T)\right)h_\alpha \Omega^\alpha\\
\delta\tau&=-\tfrac 13\rho T^2\left(\tfrac{\p\mu}{\p\vep}\right)_\rho \p_T \tfrac {a_3}T\vep^{\alpha\beta\gamma\delta} u_\alpha H_{\beta\gamma\delta}
+T\mu\bigg(\left(\mu\right.\\
&\left.\left.+\rho\left(\tfrac{\p\mu}{\p\rho}\right)_\vep\right) \tfrac{a_1}\mu
+\rho T^2 \left(\tfrac{\p\mu}{\p\vep}\right)_\rho \p_T( a_1+2\tfrac {a_3}T)\right) h_\alpha \Omega^\alpha\\
\ell^\mu&=a_1 T\mu \tilde\Delta^\mu_\nu \Omega^\nu\\
t^{\mu\nu}&=0\\
m^\mu&=-\tfrac{T^2}{\vep+p}\left(2\tfrac{s}\mu \p_T\tfrac {a_3}T-\rho\p_T\tfrac {a_1}T\right) s_1^\mu\\
&+2\left(\tfrac{\rho}{\vep+p}a_1-2\tfrac{sT}{\mu(\vep+p)}(a_2+\tfrac{a_3}T) \right)v_1^\mu +T a_1\Delta^\mu_\nu \mathcal B^\nu\\
s^{\mu\nu}&= 2(a_2+\p_T a_3)\vep^{\mu\nu\alpha\beta} u_\alpha h_\beta h^\gamma \p_\gamma T\\
&2\mu(T a_2+a_3)\vep^{\mu\nu\alpha\beta} u_\alpha h_\beta h^\gamma \p u_\gamma
\ .\end{split}\ee
In \cite{Hernandez:2017mch}, the charge-odd part of first order constitutive relations is
\be \ell^\mu=-\tilde\eta_\parallel \tilde \Sigma^\mu,\quad t^{\mu\nu}=-\tilde\eta_\perp \sigma\sigma^{\mu\nu}_\perp,\quad m^\alpha=-\tilde r_\perp \tilde Y^\alpha\ ,\ee
where
\be\label{compara} \begin{gathered}\tilde Y^\mu=\vep^{\mu\nu\alpha\beta} u_\nu h_\alpha\left(T \p\tfrac\mu T+ u^\gamma h^\sigma H_{\gamma\beta\sigma} -h^\gamma\nabla_\gamma (\mu h_\beta)\right)\\
\tilde\sigma^{\mu\nu}_\perp= \sigma_{\perp\gamma}^{(\mu}\vep^{\nu)\alpha\beta\gamma}u_\alpha h_\beta\\
\tilde \Sigma^\mu=\vep^{\mu\nu\rho\sigma}u_\nu h_\rho\tilde\Delta_{\sigma}^\alpha h^\beta\sigma_{\alpha\beta}
\ ,\end{gathered}\ee
and where
\be\begin{gathered} \sigma^{\mu\nu}_\perp=(\tilde \Delta^{\mu\alpha}\tilde \Delta^{\nu\beta}-\tfrac 12 \tilde \Delta^{\mu\nu}\tilde\Delta^{\alpha\beta})\sigma_{\alpha\beta}\\
\sigma^{\mu\nu}=(\Delta^{\mu\alpha}\Delta^{\nu\beta}-\tfrac 12 \Delta^{\mu\nu}\Delta^{\alpha\beta})\nabla_{(\alpha}u_{\beta)}\\
\tilde \Delta^{\mu\nu}=\Delta^{\mu\nu}-h^\mu h^\nu\ .
\end{gathered}\ee
Note the (on-shell) relation
\be v_1^\mu=-\frac{\rho\mu T}{\vep+p}\tilde Y^\mu\ ,\ee
while the other operators defined in (\ref{compara}) are independent from the first order operators used in (\ref{compar}). The comparison then gives
\be \begin{gathered}\tilde r_\perp=-2\frac{\rho\mu T}{\vep+p}\left(\tfrac{\rho}{\vep+p}a_1-2\tfrac{sT}{\mu(\vep+p)}(a_2+\tfrac{a_3}T) \right)\\
\tilde\eta_\parallel=0,\quad\tilde \eta_\perp=0\ .\end{gathered}\ee
In other words, except $\tilde r_\perp$, the first order terms of
\cite{Hernandez:2017mch} are not reproduced by the action (\ref{439}). The fact that not all non-dissipative transport is captured by the action was already
observed e.g. in 2+1 parity-breaking systems \cite{Haehl:2013kra}.\footnote{In \cite{Hernandez:2017mch} it is noted that $\tilde r_\perp$ is related to the Hall
conductivity. The fact that our action recovers $\tilde r_\perp$ is then consistent with that \cite{Haehl:2013kra} recovers the Hall conductivity (at least for
restricted values of the latter).} Finally we note that \cite{Hernandez:2017mch} did not consider the most general first order constitutive relations, which is why in (\ref{compar}) we have additional new terms.

\section{Conclusions}\label{sec:concl}
In this paper we constructed a variational principle for MHD following the approach of effective field theory. This theory extends previous variational formulations of hydrodynamics. The main new ingredients here are the implementation of the 1-form $U(1)$ symmetry and the introduction of the dual photon. We studied in detail the action for ideal MHD, and how, by enhancing the time shift symmetry (\ref{41}) to arbitrary time reparametrizations (\ref{41a}) one obtains a variational principle for FFE. We also observed that FFE can be recovered from a particular limit of axion electrodynamics. It would be interesting to further investigate further this point, which might potentially lead to new connections between low-density plasmas in astrophysics and topological insulators in condensed matter.

We then showed that certain subgroups of the symmetries imposed on the action lead to various conserved quantities, which are a generalization of local and non-local helicities discussed in the literature. Curiously, we have not yet been able to recover the conserved magnetic helicity in ideal MHD. In the traditional formulation such quantity is the integral of an expression depending on the gauge potential. The latter is non-locally related to the dual photon used here, so it could perhaps be that the magnetic helicity in this formulation cannot be written locally even using the Lagrangian variables.

Lastly, we discussed first order corrections to the action. These do not capture the full non-dissipative first order MHD. It would be interesting to understand if part of the missing terms can be described using a WZW-like construction, similar to that of \cite{Geracie:2014iva}. It is also desirable to extend this variational principle to include dissipative terms as well as anomalies, which should be possible using the formalism introduced in \cite{Crossley:2015tka}. As one further application, we hope that this approach can be used to address stability of objects where the magnetic field plays a significant role, in a similar spirit as what was done e.g. in \cite{friedman,Vasil:2013kua}.

\vspace{0.2in}   \centerline{\bf{Acknowledgements}} \vspace{0.2in}
We are grateful to Hong Liu, Maxim Lyutikov and Ellen Zweibel for insightful conversations. P. G. was supported by a Leo Kadanoff Fellowship. The work
of D.T.S. is supported, in part, by US Department Of Energy grant No. DE-FG02-13ER41958 and a Simons Investigator Grant from the Simons Foundation.  While this work was being completed, we learned of \cite{armc}, which has some overlap with our manuscript.

\appendix

\section{Useful formulae}\label{app:for}
\subsection{Variations with respect to background sources}
Introduce
\be\label{lam}\lambda_i^\mu=\frac{\p x^\mu}{\p \sigma^i}+v_i \frac{\p x^\mu}{\p \sigma^0}=d_i x^\mu\ .\ee
Variations with respect to the background are given by
\be\begin{gathered}\delta_g T=\frac T2 u^\mu u^\nu,\quad \delta_g v_i = T u^{(\mu} \lambda_i^{\nu)},\quad \delta_g m_i=\frac 12 u^\mu u^\nu m_i \\
\delta_g \mu=\frac \mu2(u^\mu u^\nu - h^\nu h^\mu),\quad
\delta_g\mu_N = \frac{\mu_N}2 u^\mu u^\nu \\
\delta_g a^{ij}=-\lambda^{i(\mu}\lambda^{j\nu)},\quad \delta_g m_{ij}=-2m_{[i}\lambda_{j]}^{(\nu} u^{\mu)} \\
\delta_g b_i=\mu_N u^{(\mu}\lambda_i^{\nu)},\quad \delta_g \vep^{ijk}=-\frac 12 \Delta^{\mu\nu}\vep^{ijk}\\
\delta_b m_{ij}=\lambda_i^{[\mu}\lambda_j^{\nu]},\quad \delta_b m_i=u^{[\mu}\lambda_i^{\nu]}\\
\delta_C \mu_N=\frac 12u^\mu,\quad \delta_C b_i=\lambda_i^\mu\ ,
\end{gathered}\ee
where $\delta_g$ stands for the variation with respect to $g_{\mu\nu}$, and similarly for $\delta_b$ and $\delta_C$.

\subsection{Pull-backs}
From (\ref{vel}) and (\ref{lam}), and using $\p_i\p_0 x^\mu=\p_0\p_i x^\mu$ and $\p_i\p_j x^\mu=\p_j\p_i x^\mu$, we find
\be\begin{split}
T\p_0 v_i&=\lambda_i^\mu(\p_\mu T+T\p u_\mu)\\
T\p_0\lambda_i^\mu&=\lambda_i^\nu(\nabla_\nu u^\mu+u^\mu\p u_\nu)\ ,
\end{split}\ee
from which we obtain
\be\begin{split}
T\p_0 m_{ij}&=\lambda_i^\alpha\lambda_j^\beta( u^\gamma H_{\gamma\alpha\beta}+2\nabla_{[\alpha}(\mu h_{\beta]})-2\mu h_{[\alpha}\p u_{\beta]})\\
d_{[i}v_{j]}&=T\lambda_i^\alpha\lambda_j^\beta\p_{[\alpha}u_{\beta]}\\
d_{[i}m_{j]}&=\lambda_i^\alpha\lambda_j^\beta\p_{[\alpha}(\mu h_{\beta]})\\
d_{[i}m_{jk]}&=\lambda_{[i}^\mu\lambda_j^\nu\lambda_{k]}^\rho(\tfrac 13 H_{\mu\nu\rho}+2 \mu h_\mu \nabla_\nu u_\rho)\\
T\p_0 b_i&=\lambda_i^\alpha(\p_\alpha\mu_N+\mu_N\p u_\alpha-u^\gamma D_{\alpha\gamma})\\
d_{[i}b_{j]}&=\lambda_i^\alpha\lambda_j^\beta (D_{\alpha\beta}+2\mu_N\p_{[\alpha}u_{\beta]})\\
\end{split}\ee

\subsection{Ideal equations of motion}
In this section we give the explicit form of the ideal equations of motion, which are necessary to find the on-shell independent tensor structures used in the main text. The ideal equations of motion are obtained by plugging (\ref{413}) and (\ref{JJ}) in
\be \nabla_\mu T^{\mu\nu}=\frac 12 H_{\mu\alpha\beta}J^{\alpha\beta}+D^{\nu\alpha}J_\alpha,\quad \nabla_\mu J^{\mu\nu}=0,\quad \nabla_\mu J^\mu=0\ .\ee
In the scalar sector we have five equations. Three of them are
\be\begin{split} u_\mu \nabla_\nu T^{\mu\nu}&=-\p\vep-(\vep+p)\theta+\mu\rho h^\mu h^\nu \nabla_\mu u_\nu=0\\
\mu h_\mu \nabla_\nu J^{\nu\mu}&=\mu\p\rho+\mu\rho\theta-\rho\mu h^\mu h^\nu\nabla_\mu u_\nu=0\\
\mu_N \nabla_\mu J^\mu&=\mu_N \p n+n\mu_N\theta=0\ .\end{split}\ee
Summing them, we find
\be -\p\vep+\mu_N \p n+\mu\p\rho+(\mu_N n+\mu\rho-\vep-p)\theta=-T\nabla_\mu (su^\mu),\ee
where we used
\be\label{1stl} \vep+p=sT+\mu \rho+\mu_N n,\quad d\vep=Tds+\mu d\rho+\mu_N dn.\ee
The other two scalar equations are
\be\begin{split} h_\mu \nabla_\nu T^{\mu\nu}=&(\vep+p)h^\mu \p u_\mu+\tfrac{\vep+p-\mu\rho-\mu_N n}T h^\mu \p_\mu T\\
&-\mu \nabla_\nu(\rho h^\nu)=n h^\nu\left(u^\alpha D_{\nu\alpha}-T\p_\nu\tfrac{\mu_N}T\right) \\
u_\mu \nabla_\nu J^{\nu\mu}=&\nabla_\mu(\rho h^\mu)-\rho h^\mu \p u_\mu=0,\end{split}\ee
where the second equation is a constraint. Rearranging these two equations gives
\be\begin{gathered} \nabla_\mu(\rho h^\mu)=-\tfrac\rho T h^\mu \p_\mu T\\
h^\mu\left(\p u_\mu+\tfrac{\p_\mu T}T\right)=\tfrac n{\vep+p-\mu \rho} h^\nu\left(u^\alpha D_{\nu\alpha}-T\p_\nu\tfrac{\mu_N}T\right)\ .\end{gathered}\ee
For the vector sector we have
\be\begin{split}
\tilde \Delta_{\mu\alpha} \nabla_\nu J^{\nu\alpha}=&\rho\tilde\Delta_\mu^\nu(\p h_\nu - h^\alpha \nabla_\alpha u_\nu)=0\\
\tilde\Delta_{\mu\alpha}\nabla_\nu T^{\nu\alpha}=&\tilde \Delta_\mu^\nu\bigg[\frac{\vep+p}T \left(T\p u_\nu+ \p_\nu T\right)+ nT\p_\nu \frac{\mu_N}T\\
&+\rho T \p_\nu\frac \mu T-\mu\rho h^\alpha\nabla_\alpha h_\nu\bigg]\\
=&\frac 12\rho u^\alpha h^\beta H_{\mu\alpha\beta} +n D_{\mu\nu} u^\nu+nh_\mu h^\alpha u^\nu D_{\alpha\nu},
\end{split}\ee
where $\tilde\Delta_{\mu\nu}=\Delta_{\mu\nu}-h_\mu h_\nu$. Using the tensor structures (\ref{ts1}) and (\ref{tsa1}), the second equation above can be written as
\be\label{vece}\begin{split}& s v_1^\mu+n v_2^\mu\\
&=\frac{\rho\mu}2 \tilde \Delta^\mu_\nu \varepsilon^{\nu\alpha\beta\gamma} u_\alpha (\tfrac 12 u^\rho H_{\rho\beta\gamma}+2 \p_\beta (\mu h_\gamma)-2 \mu h_\beta\p u_\gamma)\ .\end{split}\ee

\section{First order MHD with number conservation}\label{app:1mhd}
The first order action (\ref{439}) admits additional terms when including a conserved number current. The most general first order action which is linear in the electromagnetic fields is
\be\begin{split} S_{(1)}=&\int d^4 x\sqrt{-g}\big( a_1 \vep^{ijk} m_i d_j v_k\\
&+a_2T\vep^{ijk}\p_0 m_{ij} v_k+a_3 \vep^{ijk}d_i m_{jk}\\
&+a_4\vep^{ijk}m_i d_j b_k+a_5 T\vep^{ijk}\p_0 m_{ij}b_k\big)\ ,\end{split}\ee
where $a_1,a_3,a_4$ are generic functions of $T,\mu_N$. As discussed around (\ref{act1}), invariance under time shifts (\ref{41}) forces $a_2$ to be constant, and in a similar way one sees that (\ref{cs}) imposes constancy of $a_5$. Assuming that the number density $n$ does not change sign under neither of time-reversal, parity and charge-conjugation, the behavior under these transformations of $\mu_N$ and $b_i$ is the same as that of $T$ and $v_i$, respectively, which implies that the above action preserves parity and time-reversal, and breaks charge-conjugation. The coefficients $a_1,a_3,a_4$ are of thermodynamic nature as they survive in the static limit. To the constitutive relations (\ref{const1}) we have additional contributions:
\begin{widetext}
\be\begin{split}
\delta\vep&=\mu_N\mu\p_{\mu_N}(a_1+a_3)h_\mu \Omega^\mu-\frac {\mu_N}{12} \p_{\mu_N}a_3\vep^{\alpha\beta\gamma\delta}u_\alpha H_{\beta\gamma\delta} +\frac \mu2(T\p_T a_4+\mu_N\p_{\mu_N}a_4)h_\mu N^\mu\\
q^\mu&=-(T\p_{\mu_N}a_1-2\p_{\mu_N}a_3+\mu_N\p_{\mu_N} a_4)s_2^\mu-\mu_N T\p_T \tfrac {a_4}Ts_1^\mu+a_4\mu_N\mathcal B^\mu -\mu_N(\tfrac{a_4}T-\tfrac{2s}{\rho\mu}a_5)v_1^\mu \\
&-(a_4+2(1-\tfrac{\mu_N n}{\rho\mu})a_5+(a_2T+a_3)\tfrac{2 n}{\rho\mu})v_2^\mu+2\mu_Na_5 \mathcal B^\alpha h_\alpha h^\mu-\tfrac{\mu_N}{6}a_5\vep^{\alpha\beta\gamma\delta}h_\alpha H_{\beta\gamma\delta}h^\mu \\
\kappa^\mu&=\frac 12 a_4 N^\mu\\
j^{\mu\nu}&=2a_5\vep^{\mu\nu\alpha\beta}u_\alpha(\p_\beta\mu_N+\mu_N\p u_\beta-D_{\beta\delta}u^\delta)+2\p_{\mu_N}a_3\vep^{\mu\nu\alpha\beta} u_\alpha\p_\beta\mu_N
\end{split}\ee
\end{widetext}
and the additional first order constitutive relation for the number current:
\vspace{-0.6cm}

\be J_{(1)}^\mu= \delta nu^\mu+j^\mu\ ,\ee
where $j^\mu$ is transverse to $u^\mu$, and
\be\begin{split}
\delta n&=\mu(\tfrac 12T \p_{\mu_N}a_1+\p_{\mu_N}a_3)h_\alpha \Omega^\alpha\\
&-\frac 1{12} \p_{\mu_N}a_3 \vep^{\alpha\beta\gamma\delta}u_\alpha H_{\beta\gamma\delta}\end{split}\ee

\be\begin{split}
j^\mu&=-(\p_T a_4-\tfrac {a_4}T)s_1^\mu-\p_{\mu_N}a_3 s_2^\mu\\
&+a_3\mathcal B^\mu -(\tfrac {a_3}T -\tfrac{2s}{\rho\mu}a_4)v_1^\mu+\frac{2 n}{\rho\mu}a_4 v_2^\mu \\
&+2a_4\mathcal B^\alpha h_\alpha h^\mu-\frac{a_4}{6}\vep^{\alpha\beta\gamma\delta}h_\alpha H_{\beta\gamma\delta} h^\mu\ ,
\end{split}\ee
and where
\be\begin{gathered} \label{tsa1} N^\mu=\vep^{\mu\alpha\beta\gamma}u_\alpha(D_{\beta\gamma}+2\mu_N\p_\beta u_\gamma)\\ v_2^\mu=\mu\varepsilon^{\mu\alpha\beta\delta}u_\alpha h_\beta(\p_\delta \mu_N+\mu_N\p u_\delta-D_{\delta\sigma} u^\sigma)\\
s_2^\mu=\mu\varepsilon^{\mu\alpha\beta\gamma} u_\alpha h_\beta \p_\gamma \mu_N
\end{gathered}\ee

\end{document}